\documentclass[aps,twocolumn,pra,superscriptaddress,showpacs,tightenlines]{revtex4}
\usepackage{amssymb}
\usepackage{amsmath}
\usepackage{graphicx}
\usepackage{epsfig}
\usepackage{txfonts}
\usepackage{subfigure}
\usepackage{amsfonts}
\usepackage{CJK}

\begin{document}
\title{Tailorable couplings of a cantilever with a superconducting charge qubit:\\ Quantum state engineering}
\author{Jie-Qiao Liao}
\affiliation{Key Laboratory of Low-Dimensional Quantum Structures
and Quantum Control of Ministry of Education, and Department of
Physics, Hunan Normal University, Changsha 410081, China}
\author{Le-Man Kuang\footnote{Corresponding author}}
\affiliation{Key Laboratory of Low-Dimensional Quantum Structures
and Quantum Control of Ministry of Education, and Department of
Physics, Hunan Normal University, Changsha 410081, China}

\begin{abstract}
We propose a theoretical scheme to realize tailorable couplings
between a cantilever and a superconducting charge qubit. By tuning
the controllable parameters of the qubit, both linear and nonlinear
couplings between the cantilever and the qubit can be achieved.
Based on these couplings, we show the preparation of the cantilever
into some interesting quantum states, such as superposed coherent
states and squeezed states, via manipulating and detecting the
qubit. We also study the influence of the environment on quantum
states of the cantilever. It is indicated that decoherence induced
by the environment can drive the cantilever from  superposed
coherent states  into the steady coherent state. It is also found
that the environment can induce the steady-state position squeezing
of the cantilever under a critical temperature. These results will
shed new light on production of nonclassical effects of the
cantilever.
\end{abstract}


%85.85.+j Micro- and nano-electromechanical systems (MEMS/NEMS) and devices
%85.25.Cp Josephson devices
%42.50.Dv quantum state engineering and measurements

\maketitle

\section{\label{introduction}Introduction}

As is well known, it is of very significance to couple a mechanical
object to an electronic system  since the electronics may be used to
measure the quantum nature of the mechanical object while quantum
effects in electronic systems may be measured by mechanics.
Cantilevers, as important components of magnetic resonance force
microscopy (MRFM)
 \cite{Sidles1995,Mamin2003,Bermangroup}, have attracted much
attention of both theorists and experimentalists in many fields of
physics such as condensed matter physics and quantum information
science \cite{Cleland2003,Blencowe2004,Schwab2005}. For example,
MRFM has been recently proposed as a qubit readout device for
spin-based quantum computers~\cite{DiVincenzo1995,Berman2000}. As
one kind of nanomechanical resonators, the cantilever can also be
used as a platform to study some fundamental problems in quantum
mechanics such as quantum measurement~\cite{Schwab2004}, quantum
decoherence~\cite{Zurek1991}, the boundary between quantum and
classical communities~\cite{Katz2007}, and test of quantum mechanics
in macroscopic scales~\cite{Armour2002,Bouwmeester2003}. By far,
with the high-speed development of modern micro-fabrication
techniques, the preparation of nanomechanical resonators with high
frequency and high quality factor
~\cite{Knobel2003,Huang2003,Gaidarzhy2005} has become possible. At
the same time, efficient cooling technologies
~\cite{Imamoglu2004,Karrai2004,Zhang2005,Naik2006,Zeilinger2006,Heidmann2006,Bouwmeester2006,Poggio2007,Wilson2007,Marquardt2007,Xue2007B,Yi2008,Wang2009,Ouyang2009}
drive nanomechanical resonators approaching quantum realm.
Therefore, it is expected to observe quantum
evidences~\cite{Quantumevidence} in nanomechanical resonators such
as cantilevers. As a precondition, it is an interesting topic that
how to generate some nonclassical states ~\cite{
Xue2007A,Liao2008,Jacobs2007,Siewert2005,Tian2005,Rabl2004,Ruskov2005,Zhou2006,Zhang2009,Hou2007,Buks2007,Eisert2004,Bos2005,Vitali2007}
such as Fock states, superposed states, squeezed states, and
entangled states in cantilevers. In addition, from the viewpoint of
MRFM, cantilevers play the role of probers to read out the states of
a single spin, therefore the preparation of cantilevers into some
states with position squeezing can improve the measurement
precision~\cite{Scully1997}.

For manipulation of the vibrational states of nanomechanical
resonators, many schemes have been proposed by coupling
nanomechanical resonators with various physical systems such as
superconducting qubits
~\cite{Armour2002,Irish2005,Schwab2009,Sun2006,Xue2007C},
superconducting transmission line resonators
~\cite{Blencowe2007,Rocheleau2009,Hertzberg2009}, cold ions, atoms
and molecules
~\cite{Tian2004,Treutlein2007,Treutlein2010,Meystre2010}, and
quantum dots ~\cite{Liao2008,Lambert2008,Lambert2008B,Bennett2010}.
These considered physical systems play the role of indirectly
assistant controller~\cite{Jacobs2007B}. Therefore, they should be
of well controllability and readability. For example, during the
latest decade, great advances in quantum information processing
based on superconducting charge qubits have been
made~\cite{Makhlin2001,You2005}. It has been shown that
superconducting charge qubits have well controllability and
readability~\cite{Nakamura1999}. Stimulated by these, in this paper,
we propose a new scheme to couple a cantilever with a
superconducting charge qubit~\cite{Makhlin1999}. We can manipulate
the cantilever through tuning and measuring the qubit as an indirect
controller. Both linear and nonlinear couplings~\cite{Zhou2006}
between the cantilever and the qubit can be obtained. Especially, we
emphasize that the obtained \textit{nonlinear}-type coupling for a
\textit{cantilever} is one of the main results in this work. Based
on these couplings, we show how to create superposed coherent states
and squeezed states of the cantilever by manipulation of the qubit.
The preparation of superposed coherent states shows the appearance
of quantum superposition in the cantilever. And the squeezing in the
cantilever not only shows nonclassical evidence but also has wide
potential for practical application.

This paper is organized as follows. In Sec.~\ref{physicalmodel}, we
introduce the physical model and present correspondent Hamiltonian.
In Sec.~\ref{supercoherentstate}, we show how to generate superposed
coherent states of the cantilever, and investigate the influence of
decoherence induced by environment on superposed coherent states of
the cantilever. In Sec.~\ref{dynamicsqueezing}, we show how to
obtain a dynamical squeezing of the cantilever, especially it is
indicated that there exists a steady-state position squeezing. We
conclude this paper with some discussions in the last section.
Finally, we present two Appendixes~\ref{appenA} and \ref{appendixb}
for derivation of evolution equations of the cantilever in an
environment.

\section{\label{physicalmodel}Physical model and Hamiltonian}

\begin{figure}[htp]
\center
\includegraphics[bb=49 586 442 779,width=3.2 in]{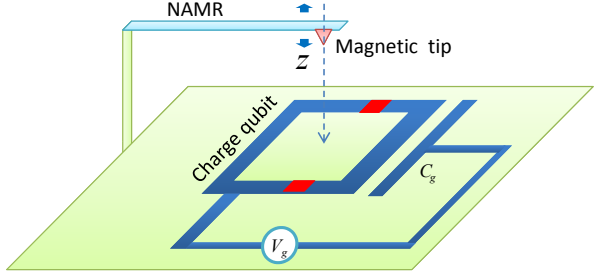}
\caption{(Color online) Setup of a cantilever mounted with a
magnetic tip coupling with an SQUID-based charge qubit. The magnetic
field generated by the magnetic tip threads through the
superconducting loop of the charge qubit. Since the generated
magnetic field depends on the vibration amplitude $z$ of the
cantilever, and the magnetic flux in the loop is a controllable
parameter of the qubit, therefore a coupling between the cantilever
and the qubit is induced.}\label{scheme}
\end{figure}

We start with introducing the physical setup as illustrated in
Fig.~\ref{scheme}, a cantilever is fabricated above a
superconducting charge qubit which is formed by an SQUID-based
Cooper-pair box~\cite{Makhlin1999}.
%%%%%%%%%%%%%%%%%%%%%%%%%%%%%%%%%%%%%%%%%%%%%%%%%%%%%%%%%%%%%%%%%%
%%%%%%%%%%%%%%%%%%%%%%%%%%%%%%%%%%%%%%%%%%%%%%%%%%%%%%%%%%%%%%%%%%
A ferromagnetic particle mounted on the cantilever tip produces a
magnetic field~\cite{Xue2007}
\begin{eqnarray}
B_{\textrm{tip}}=\frac{\mu_{0}}{4\pi
r^{3}}\left[3\left(\vec{n}\cdot\vec{m}\right)\vec{n}-\vec{m}\right],
\end{eqnarray}
threading the superconducting loop of the qubit, where $\mu_{0}$ is
the vacuum magnetic conductance, $\vec{n}$ is the unit vector
pointing to the direction from the tip to the center of the loop.
Here we assume that the magnetic tip is right on top of the center
of the loop. The vector $\vec{m}$ is the magnetic moment of the
magnetic tip pointing to the $z$ direction, and $r$ is the distance
between the tip and the center of the loop. For a tiny vibration of
the cantilever, the magnetic field threading the loop of the qubit
can be approximated as
\begin{eqnarray}
B_{\textrm{tip}}\approx B_{0}-Cz,
\end{eqnarray}
where $B_{0}=\mu_{0}m/(2\pi r^{3})$ and $C=3\mu_{0}m/(2\pi r^{4})$
with $m=|\vec{m}|$. Since the generated magnetic field depends on
the position $z$, then couplings between the cantilever and the
qubit can be induced as follows: The vibration of the cantilever
leads to a change of the magnitude for the magnetic field threading
the loop of the qubit. As a result, the corresponding magnetic flux
will be changed. At the same time, since the magnetic flux is a
controllable parameter of the qubit, therefore the vibration can
induce a coupling between the cantilever and the qubit.

As for the SQUID-based charge qubit~\cite{Makhlin1999}, its
Hamiltonian reads
\begin{eqnarray}
\label{eq:H_cpb}
H_{q}=4E_{C}\left(n_{g}-\frac{1}{2}\right)\sigma_{z}-E_{J}\cos\left(\frac{\pi\Phi_{e}}{\Phi_{0}}\right)\sigma_{x},
\end{eqnarray}
where $E_{C}=e^{2}/[2(2C_{J}+C_{g})]$ is the single-electron
charging energy, $C_{J}$ and $C_{g}$ are, respectively, the
capacitances of each Josephson junction and the gate capacitance in
the qubit; $n_{g}=C_{g}V_{g}/2e$ is the gate charge number with
$V_{g}$ being gate voltage; $E_{J}$ is the Josephson coupling
energy; $\Phi_{e}$ and $\Phi_{0}$ are the externally biasing flux
and flux quanta, respectively. The Pauli operators introduced in
Eq.~(\ref{eq:H_cpb}) are defined by
\begin{eqnarray}
\sigma_{z}&=&|0\rangle\langle0|-|1\rangle\langle1|,\hspace{0.5 cm}
\sigma_{x}=|0\rangle\langle1|+|1\rangle\langle0|,
\end{eqnarray}
where the states $|0\rangle$ and $|1\rangle$ represent that there is
no and one extra cooper pair on the inland, respectively.

We consider a case that the externally biasing magnetic flux
$\Phi_{e}$ is composed of two parts. One is generated by the
magnetic tip, and the other is generated by an externally
controllable electric current. We can express the total biasing
magnetic flux as $\Phi_{e}=(B_{\textrm{tip}}+B_{x})S$, where $B_{x}$
is the magnetic field generated by the externally controllable
electric current, $S$ is the area of the superconduting loop. For a
tiny vibration, the cantilever can be modeled as a quantum harmonic
oscillator, which is depicted by the usual Bosonic creation and
annihilation operators $a^{\dagger}$ and $a$, satisfying the
commutative relation $[a,a^{\dagger}]=1$. Then the magnetic field
generated by the magnetic tip can be expressed as
\begin{eqnarray}
B_{\textrm{tip}}\approx B_{0}-Cz_{0}(a^{\dagger}+a),
\end{eqnarray}
where $z_{0}=1/\sqrt{2\mu\omega}$ (with $\hbar=1$) is zero-point
uncertainty for the ground state of the cantilever, $\mu$ and
$\omega$ are the mass and frequency of the cantilever, respectively.

The Hamiltonian of the total system including the cantilever and the
charge qubit reads
\begin{eqnarray}
\label{eq:H_tot} H=\frac{\omega_{0}}{2}\sigma_{z}+\omega
a^{\dagger}a
-E_{J}\cos[\phi_{0}+\phi(a^{\dagger}+a)]\sigma_{x},
\end{eqnarray}
where $\omega_{0}=8E_{C}(n_{g}-1/2)$ and $\omega$ are, respectively,
the frequencies of the qubit and the cantilever. We also introduce
two parameters
\begin{eqnarray}
\phi_{0}=\pi S(B_{0}+B_{x})/\Phi_{0},\hspace{0.5 cm} \phi=-\pi
SCz_{0}/\Phi_{0}.
\end{eqnarray}
Hamiltonian~(\ref{eq:H_tot}) obviously shows a nonlinear coupling
between the cantilever and the charge qubit. A coupling of similar
form between a bosonic mode and a two-level system has been obtained
in a trapped-ion system~\cite{Monroe2005}. A recent scheme has been
proposed to obtain a nonlinear interaction between a doubly-clamped
beam and a superconducting charge qubit~\cite{Zhou2006}. However,
the method proposed in Ref.~\cite{Zhou2006} is not valid for a
cantilever.

Hamiltonian~(\ref{eq:H_tot}) is very useful in quantum information
processing. Many useful interactions can be tailored from
Eq.~(\ref{eq:H_tot}) by choosing proper parameters. For example, we
tune the externally controllable current such that $\cos\phi_{0}=0$,
then Eq.~(\ref{eq:H_tot}) becomes
\begin{eqnarray}
\label{e3} H_{1}=\omega a^{\dagger}a+\frac{\omega_{0}}{2}\sigma_{z}
+E_{J}\sin[\phi(a^{\dagger}+a)]\sigma_{x},
\end{eqnarray}
which can be reduced to the well-known Jaynes-Cummings Hamiltonian
without rotating wave approximation
\begin{eqnarray}
\label{e4} H_{1}\approx\frac{\omega_{0}}{2}\sigma_{z}+\omega
a^{\dagger}a+g(a^{\dagger}+a)\sigma_{x},
\end{eqnarray}
by expanding the sine function up to the first order of parameter
$\phi$, where we introduce the coupling strength $g=E_{J}\phi$.

On the other hand, if we choose the external magnetic flux to ensure
$\cos\phi_{0}=1$, then Eq.~(\ref{eq:H_tot}) reduces to
\begin{eqnarray}
\label{e5} H_{2}=\omega a^{\dagger}a+\frac{\omega_{0}}{2}\sigma_{z}
-E_{J}\cos[\phi(a^{\dagger}+a)]\sigma_{x},
\end{eqnarray}
which can be further simplified by expanding the cosine function up
to the second order of $\phi$,
\begin{eqnarray}
\label{e6} H_{2}\approx\omega
a^{\dagger}a+\frac{\omega_{0}}{2}\sigma_{z}-E_{J}\sigma_{x}
-g'(a^{\dagger}+a)^{2}\sigma_{x},
\end{eqnarray}
where $g'=E_{J}\phi^{2}/2$ is a nonlinear coupling strength between
the cantilever and the qubit. These two kinds of couplings given in
Eqs.~(\ref{e4}) and (\ref{e6}) are very useful in quantum optics and
quantum information processing. As examples, in the following two
sections, we will study quantum state engineering based on these
couplings.

\section{\label{supercoherentstate}Preparation of superposed coherent states}

Superposed coherent states are typical quantum states which exhibit
nonclassical properties. In this section, we show how to prepare the
cantilever into superposed coherent states with the above obtained
Hamiltonian~(\ref{e4}). We also study the decoherence of the created
superposed coherent states when the cantilever is subjected to an
environment.

\subsection{Generation of superposed coherent state}

Firstly, we consider an ideal situation in which there is no
dissipation for the cantilever. Since the state preparation can be
realized in a very short time interval, it is reasonable to neglect
the dissipation during the state preparation process. We tune the
gate voltage $V_{g}$ such that $n_{g}=1/2$, that is $\omega_{0}=0$,
then Eq.~(\ref{e4}) reduces to the conditional displacement harmonic
oscillator (CDHO) Hamiltonian
\begin{eqnarray}
\label{e7} H_{\textrm{CDHO}}=\omega
a^{\dagger}a+g(a^{\dagger}+a)\sigma_{x}.
\end{eqnarray}
Corresponding to the qubit in states $|\pm\rangle$, the displacement
terms are $\pm g(a^{\dagger}+a)$, respectively, where states
$|\pm\rangle$ are the eigenstates of Pauli operator $\sigma_{x}$,
with the respective eigenvalues $\pm1$.

For generation of superposed coherent states, we suppose the total
system consisting of the cantilever and the qubit is initially
prepared in a state
$|\varphi(0)\rangle=|\alpha_{i}\rangle\otimes|0\rangle_{q}$, where
$|\alpha_{i}\rangle$ is the usual Glauber coherent state, which is
defined as the eigenstate of annihilation operator $a$, i.e.,
$a|\alpha_{i}\rangle=\alpha_{i}|\alpha_{i}\rangle$, and $|0\rangle$
is defined by $|0\rangle=(|+\rangle+|-\rangle)/\sqrt{2}$. Making use
of Hamiltonian~(\ref{e7}), the state of the total system at time $t$
is
\begin{eqnarray}
\label{e8}|\varphi(t)\rangle&=&\exp\left(-itH_{\textrm{CDHO}}\right)|\varphi(0)\rangle\nonumber\\
&=&\frac{1}{2}\left[\left(e^{i\theta_{+}}|\alpha_{+}\rangle+e^{i\theta_{-}}|\alpha_{-}\rangle\right)|0\rangle\right.\nonumber\\
&&\left.+\left(e^{i\theta_{+}}|\alpha_{+}\rangle-e^{i\theta_{-}}|\alpha_{-}\rangle\right)|1\rangle\right],
\end{eqnarray}
where we introduce the parameters
\begin{subequations}
\label{e9}
\begin{align}
\alpha_{\pm}&=\alpha_{i} e^{-i\omega t}\pm \frac{g}{\omega}(e^{-i\omega t}-1),\\
\theta_{\pm}&=\frac{g}{\omega}\left[gt-\left(\frac{g}{\omega}\pm\alpha_{i}\right)\sin(\omega
t)\right].
\end{align}
\end{subequations}

During the derivation of Eq.~(\ref{e8}), we have used the following
formula~\cite{Liao2007},
\begin{eqnarray}
\label{e10}
e^{[\theta(\beta_{1}a+\beta_{2}a^{\dag}a+\beta_{3}a^{\dag})]}
=e^{f_{1}a^{\dag}}e^{f_{2}a^{\dag}a}e^{f_{3}a}e^{f_{4}}
\end{eqnarray}
with
\begin{subequations}
\label{e11}
\begin{align}
f_{1}&=\beta_{3}\left(e^{\beta_{2}\theta}-1\right)/\beta_{2},\\
f_{2}&=\beta_{2}\theta,\\
f_{3}&=\beta_{1}\left(e^{\beta_{2}\theta}-1\right)/\beta_{2,}\\
f_{4}&=\beta_{1}\beta_{3}\left(e^{\beta_{2}\theta}-\beta_{2}\theta-1\right)/\beta_{2}^{2}.
\end{align}
\end{subequations}
The action of the operator given in Eq.~(\ref{e10}) on a coherent
state $|\alpha_{i}\rangle$ yields~\cite{Liao2007}:
\begin{eqnarray}
\label{e12}
e^{[\theta(\beta_{1}a+\beta_{2}a^{\dag}a+\beta_{3}a^{\dag})]}|\alpha_{i}\rangle=e^{\varepsilon}\left|f_{1}+\alpha_{i}
e^{f_{2}}\right\rangle,
\end{eqnarray}
where $\varepsilon$ is given by the following expression
\begin{eqnarray}\label{e13}
\varepsilon&=&f_{4}+f_{3}\alpha_{i}+\left(|\alpha_{i}
\exp(f_{2})|^{2}-|\alpha_{i}|^{2}+|f_{1}|^{2}\right.\nonumber\\
&&\left.+2\textrm{Re}[f_{1}\alpha_{i}^{*}\exp(f_{2}^{*})]\right)/2.
\end{eqnarray}

From Eq.~(\ref{e8}), it is obvious to prepare the cantilever into
superposed coherent states through measuring the qubit.
Corresponding to the states $|0\rangle$ and $|1\rangle$ of the qubit
are measured, the cantilever collapses to the following superposed
coherent states
\begin{eqnarray}
\label{e14}
|\psi_{\pm}\rangle=\mathcal{N}_{\pm}\left(e^{i\theta_{+}}|\alpha_{+}\rangle\pm
e^{i\theta_{-}}|\alpha_{-}\rangle\right),
\end{eqnarray}
where
$\mathcal{N}_{\pm}^{-2}=2\{1\pm\textrm{Re}[\exp(-i(\theta_{+}-\theta_{-}))\langle\alpha_{+}|\alpha_{-}\rangle]\}$
are normalization constants.

For a special case, we suppose the cantilever is initially prepared
in a vacuum state, i.e., $\alpha_{i}=0$, then the states of the
cantilever at time $t$ are the so-called Schr\"{o}dinger cat
states~\cite{Gerry}
\begin{eqnarray}
|\varphi_{\pm}\rangle=\mathcal{M}_{\pm}\left(|\beta\rangle\pm
|-\beta\rangle\right),\label{schrcatstate}
\end{eqnarray}
with $\beta=g[\exp(-i\omega t)-1]/\omega$ and
$\mathcal{M}_{\pm}^{-2}=2[1\pm \exp(-2\beta^{2})]$.

When the qubit is detected in states $|0\rangle$ or $|1\rangle$, the
coupling term $g(a^{\dagger}+a)\sigma_{x}$ between the qubit and the
cantilever will entangle them. Therefore, from the experimental
viewpoint, we should decouple the cantilever and the qubit, as long
as the superposed coherent states are prepared. The method to
decoupling is making the second measurement on the qubit in states
$|\pm\rangle$. Since states $|\pm\rangle$ are eigenstates of the
operator $\sigma_{x}$, then the qubit will stay in states
$|\pm\rangle$ forever, and the dynamics of the cantilever is
governed by the displaced harmonic oscillator (DHO) Hamiltonian
\begin{eqnarray}
H_{\textrm{DHO}}=\omega a^{\dagger}a+g_{s}(a^{\dagger}+a),
\end{eqnarray}
where $g_{s}=\pm g$ corresponding to the qubit in states
$|\pm\rangle$.

\subsection{Decoherence of superposed coherent states}

In this subsection, we investigate the decoherence of superposed
coherent states produced in the previous subsection. As a practical
physical system, the cantilever couples inevitably with its external
environment. Therefore, the cantilever prepared in superposed
coherent states will loss its coherence and energy. We suppose that
the preparation time for this initial state is very short, thus we
neglect the decoherence in the course of the initial state
preparation process.

The dynamics of the cantilever is governed by the quantum master
equation
\begin{eqnarray}
\dot{\rho}=i[\rho,H_{\textrm{DHO}}]+\mathcal{L}[\rho],\label{mastereqT}
\end{eqnarray}
where the decoherence of the cantilever is phenomenologically
represented by the superoperator $\mathcal{L}$. At a temperature of
$T$, this superoperator can be written as~\cite{Scully1997}
\begin{eqnarray}
\mathcal{L}[\rho]&=&\frac{\gamma}{2}(\bar{n}_{th}+1)(2a\rho
a^{\dagger}-a^{\dagger}a\rho -\rho a^{\dagger}a)\nonumber\\
&&+\frac{\gamma}{2}\bar{n}_{th}(2a^{\dagger}\rho
a-aa^{\dagger}\rho -\rho aa^{\dagger}),\label{dissipator}
\end{eqnarray}
where $\gamma$ is decay rate, and
$\bar{n}_{th}=1/[\exp(\omega/T)-1]$ is average thermal excitation
number of the thermal bath at frequency $\omega$.
Equation~(\ref{mastereqT}) shows that there are three kinds of
actions on the cantilever. The free Hamiltonian $\omega
a^{\dagger}a$ rotates the system in phase space. The driving term
$g_{s}(a^{\dagger}+a)$ displaces the cantilever in phase space. And
the dissipation term $\mathcal{L}$ decreases the coherence and
energy of the system.

To see the evolution of the cantilever, we need to solve quantum
master equation~(\ref{mastereqT}). In order to do this, we introduce
the following transform,
\begin{subequations}
\label{transforms}
\begin{align}
\rho^{(2)}(t)&=D^{\dagger}(\alpha(t))R^{\dagger}(\theta(t))\rho(t)
R(\theta(t))D(\alpha(t)),\\
R(\theta(t))&=\exp(-i\theta(t) a^{\dagger }a),\\
D(\alpha(t))&=\exp(\alpha(t) a^{\dagger }-\alpha^{\ast }(t) a),
\end{align}
\end{subequations}
with
\begin{subequations}
\begin{align}
\theta(t) &=\omega t,\\
\alpha(t)
&=\frac{ig_{s}}{\frac{\gamma}{2}+i\omega}\left(e^{-\frac{\gamma
t}{2}}-e^{i\omega t}\right),
\end{align}
\end{subequations}
then quantum master equation~(\ref{mastereqT}) can be transformed to
a standard form
\begin{eqnarray}
\dot{\rho}^{(2)} &=&\frac{\gamma
}{2}(\bar{n}+1)(2a\rho^{(2)}a^{\dagger }-a^{\dagger
}a\rho^{(2)} -\rho
_{s}^{(2)}a^{\dagger }a)\nonumber\\
&&+\frac{\gamma }{2}\bar{n}(2a^{\dagger }\rho^{(2)}
a-aa^{\dagger
}\rho^{(2)}-\rho^{(2)}aa^{\dagger}),\label{transformedmastereq}
\end{eqnarray}
which describes the evolution for a harmonic oscillator in a heat
bath. The detailed derivation of quantum master
equation~(\ref{transformedmastereq}) will be presented in
Appendix~\ref{appenA}.

The operators $R(\theta)$ and $D(\alpha)$ in Eq.~(\ref{transforms})
are, respectively, the usual rotation and displacement operators for
a harmonic oscillator in phase space. In principle, the solution for
quantum master equation~(\ref{transformedmastereq}) can be obtained
with the superoperator method. However, for simplicity, we only give
the analytical solutions for the zero temperature case in the
following.

Based on the above discussions, the dynamical evolution of the
cantilever can be obtained as follows: When the cantilever is
initially prepared in an initial state $\rho(0)$, then the state
$\rho(t)$ of the cantilever at time $t$ can be obtained through the
processes
\begin{eqnarray}
\rho(0)\rightarrow\rho^{(2)}(0)\rightarrow\rho^{(2)}(t)\rightarrow\rho(t).
\end{eqnarray}
The two processes $\rho(0)\rightarrow\rho^{(2)}(0)$ and
$\rho^{(2)}(t)\rightarrow\rho(t)$ are determined by the transform
given in Eq.~(\ref{transforms}) and its inverse transform, and the
evolution process $\rho^{(2)}(0)\rightarrow\rho^{(2)}(t)$ is
governed by quantum master equation~(\ref{transformedmastereq}) in
the transformed representation.

We assume that the initial state of the cantilever is
\begin{equation}
\left\vert \psi \left( 0\right) \right\rangle
=\mathcal{N}_{\varphi}\left( \left\vert \beta \right\rangle
+e^{i\varphi }\left\vert -\beta \right\rangle
\right),\label{initialstate}
\end{equation}
where $\mathcal{N}_{\varphi}=[2(1+\exp(-2\beta ^{2}) \cos \left(
\varphi \right))]^{-1/2}$ is the normalization constant. Then the
state of the cantilever at time $t$ is
\begin{eqnarray}
\rho(t)&=&\mathcal{N}_{\varphi}^{2}\left(\vert\beta_{+}\rangle\langle
\beta _{+}\vert +\vert\beta_{-}\rangle\langle \beta _{-}\vert
+e^{-i\varphi }\exp(\Delta _{r}+i\Delta
_{i})\vert\beta_{+}\rangle\langle \beta
_{-}\vert\right.\nonumber\\
&&\left.+e^{i\varphi }\exp(\Delta _{r}-i\Delta _{i})\vert
\beta_{-}\rangle \langle \beta _{+}\vert \right),\label{stateatt}
\end{eqnarray}
where the parameters are defined as
\begin{subequations}
\begin{align}
\Delta_{r}&=2\vert\beta\vert ^{2}(e^{-\gamma t}-1),\\
\Delta_{i}&=2\textrm{Im}\left( \alpha(t) \beta ^{\ast }e^{-
\frac{\gamma t}{2}}\right),\\
\beta_{\pm} &=\left(\alpha(t)\pm\beta e^{-\frac{\gamma t}{2}}\right)
e^{-i\omega t}.
\end{align}
\end{subequations}

The detailed derivation from state~(\ref{initialstate}) to
state~(\ref{stateatt}) will be given in Appendix~\ref{appendixb}.
From Eq.~(\ref{stateatt}), we can see that
$\beta_{+}(\infty)=\beta_{-}(\infty)=-ig_{s}/(\gamma/2+i\omega)$ at
the long time limit. Therefore the steady state of the cantilever is
a coherent state $\left|-ig_{s}/(\gamma/2+i\omega)\right\rangle$,
which is resulted from the net actions of the free evolution, the
coherent driving, and the decoherence. When $g_{s}=0$, the steady
state reduces to vacuum state $|0\rangle$, which implies the
cantilever approaching an equilibrium with the zero temperature
environment.

For seeing clearly the decoherence of the generated superposed
coherent states, we address the evolution of the Wigner function for
these states. The definition of the Wigner function~\cite{Burnett}
of a density operator $\rho$ is
\begin{equation}
W(\xi)=2\textrm{Tr}\left[D(-\xi)\rho D(\xi )e^{i\pi a^{\dagger
}a}\right].
\end{equation}
For the state given in Eq.~(\ref{stateatt}), the Wigner function is
obtained as
\begin{eqnarray}
W(\xi)&=&2\mathcal{N}_{\varphi}^{2}\left(\exp\left(-2\vert\xi-\beta_{+}\vert^{2}\right)
+\exp\left(-2\vert\xi-\beta_{-}\vert^{2}\right)\right.\nonumber\\
&&\left.+2\textrm{Re}\left[e^{-i\varphi
}\exp(\Delta_{r}+i\Delta_{i}+\Theta)\right]\right),\label{wingerfunction}
\end{eqnarray}
where $\Theta$ is given by the following expression
\begin{eqnarray}
\Theta=2\left(\xi\beta_{-}^{\ast}+\xi^{\ast}\beta_{+}-|\xi|^{2}\right)
-\frac{1}{2}\left(|\beta _{+}|^{2}+|\beta _{-}|^{2}+2\beta
_{-}^{\ast }\beta _{+}\right).
\end{eqnarray}
%%%%%%%%%%%%%%%%%%%%%%%%%%%%%%%%%%%%%%%%%%%%%%%%%%%%%%%%%%%%%%%%%%
\begin{figure*}[htp]
\center
\includegraphics[bb=60 177 543 591,width=6.4 in]{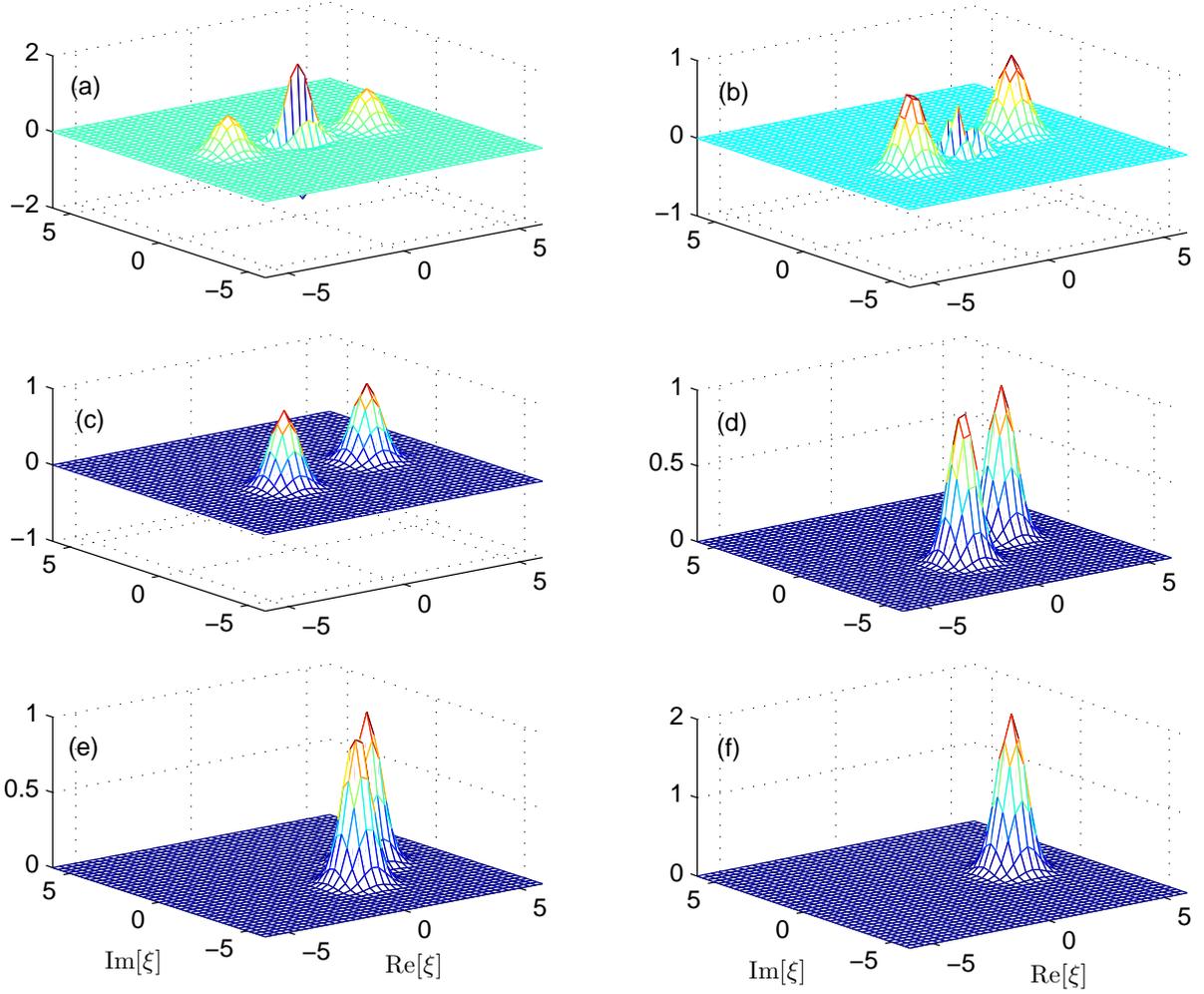}
\caption{(Color online) Plot of the Wigner function $W(\xi)$ given
in Eq.~(\ref{wingerfunction}) for the state of the cantilever at
different scaled times $\tau=\gamma t$: (a) $\tau=0$, (b)
$\tau=0.06$, (c) $\tau=0.5$, (d) $\tau=0.1$, (e) $\tau=1.5$, (f)
$\tau=20$. Other parameters are set as $\omega/\gamma=100$,
$g/\gamma=-300$, $\beta=3$, and $\varphi=0$.}\label{wingerfun}
\end{figure*}
%%%%%%%%%%%%%%%%%%%%%%%%%%%%%%%%%%%%%%%%%%%%%%%%%%%%%%%%%%%%%%%%%%
In Fig.~\ref{wingerfun}, we plot the Wigner function given in
Eq.~(\ref{wingerfunction}) at different times for the case of
$\varphi=0$ and $g_{s}=g$. When $t=0$, the Wigner function is right
for the state $|\varphi_{+}\rangle$ given in
Eq.~(\ref{schrcatstate}) with $\beta=3$. This state is the so-called
even Schr\"{o}dinger cat state. We can see from
Fig.~\ref{wingerfun}(a) that there is some coherence in
state~(\ref{schrcatstate}). This is because the Wigner function in
Fig.~\ref{wingerfun}(a) exhibits some interference fringes (namely
some oscillations), and these oscillations imply quantum coherence
in the state. With the increase of the time $t$, we can find that
the oscillations decrease gradually. At the same time, the positions
of the two main peaks of the Wigner function rotate on the phase
plain and move gradually to a point. This point corresponds
approximately to the steady state of quantum master
equation~(\ref{mastereqT}) for $T=0$. In fact, this steady state is
a coherent state
$|-ig_{s}/(\gamma/2+i\omega)\rangle\approx|2.999+0.015i\rangle$,
which is obtained from Eq.~(\ref{stateatt}) by taking the long time
limit. Moreover, with the dissipative evolution, the negative values
of the Wigner function will disappear gradually, which implies that
the nonclassical properties of the cantilever decreases with the
decoherence. Therefore, the actions of the environment and the
driving force will destroy the coherence of the superposed coherent
states and drive the cantilever into a steady coherent state.

\section{\label{dynamicsqueezing}Dynamical squeezing}

In the above section, we have study the creation of superposed
coherent states based on the obtained linear Hamiltonian. In this
section, we study the creation of dynamical squeezing as an
application of the nonlinear Hamiltonian. The created squeezed state
not only exhibits nonclassical properties, but also is useful for
precise measurement.

\subsection{Dynamical squeezing without dissipation}

From Hamiltonian (\ref{e6}), we control the gate voltage $V_{g}$
such that $\omega_{0}=0$, then Hamiltonian (\ref{e6}) becomes
\begin{eqnarray}
\label{e16}H=\omega
a^{\dagger}a-\left[E_{J}+g'(a^{\dagger}+a)^{2}\right]\sigma_{x}.
\end{eqnarray}
It can be seen from Hamiltonian~(\ref{e16}) that if the qubit is
initially prepared in one of the two eigenstates $|\pm\rangle$ of
$\sigma_{x}$, then the qubit will stay in this state forever.
Corresponding to the cases of the qubit in $|\pm\rangle$, the
conditional Hamiltonians of the cantilever are
\begin{eqnarray}
\label{e17}H_{k}=\omega_{k}a^{\dagger}a+g_{k}\left(a^{\dagger2}+a^{2}\right),\hspace{0.5
cm}k=\pm1,
\end{eqnarray}
where $\omega_{k}=\omega+2g_{k}$ and $g_{k}=-k\times g'$.

To diagonalize Hamiltonian~(\ref{e17}), we introduce a unitary
operator~\cite{Wagner1986}
\begin{eqnarray}
\exp(-S_{k})=\exp\left[-\lambda_{k}\left(a^{2}-a^{\dag2}\right)\right].\label{defitransf}
\end{eqnarray}
Using the commutative relations $[a,S_{k}]=-2\lambda_{k} a^{\dag}$
and $[a^{\dag},S_{k}]=-2\lambda_{k} a$, we obtain
\begin{subequations}
\begin{align}
e^{-S_{k}}ae^{S_{k}}&=a\cosh(2\lambda_{k})-a^{\dag}\sinh(2\lambda_{k}),\\
e^{-S_{k}}a^{\dag}e^{S_{k}}&=a^{\dag}\cosh(2\lambda_{k})-a\sinh(2\lambda_{k}).
\end{align}
\end{subequations}
Application of the transform defined in Eq.~(\ref{defitransf}) on
Hamiltonian~(\ref{e17}) leads to
\begin{eqnarray}
\label{e20}\tilde{H}_{k}&\equiv&\exp(-S_{k})H_{k}\exp(S_{k})\nonumber\\
&=&\Omega_{k}a^{\dag}a+\Lambda_{k}\left(a^{\dag2}+a^{2}\right)+C_{k},
\end{eqnarray}
where we introduce the following parameters
\begin{subequations}
\begin{align}
\Omega_{k}&=\omega_{k}\cosh(4\lambda_{k})-2g_{k}\sinh(4\lambda_{k}),\\
\Lambda_{k}&=-\omega_{k}\sinh(4\lambda_{k})/2+g_{k}\cosh(4\lambda_{k}),\\
C_{k}&=\omega_{k}\sinh^{2}(2\lambda_{k})-g_{k}\sinh(4\lambda_{k}).
\end{align}
\end{subequations}
By choosing proper parameter $\lambda_{k}$ to ensure
$\Lambda_{k}=0$, namely $\tanh(4\lambda_{k})=2g_{k}/\omega_{k}$,
then we may choose
\begin{eqnarray}
\sinh(4\lambda_{k})&=&\frac{2g_{k}}{\Omega_{k}},\hspace{0.5 cm}
\cosh(4\lambda_{k})=\frac{\omega_{k}}{\Omega_{k}},
\end{eqnarray}
where $\Omega_{k}=\sqrt{\omega_{k}^{2}-4g_{k}^{2}}$. Therefore we
obtain the diagonalized Hamiltonian
\begin{eqnarray}
\label{e24}\tilde{H}_{k}=\Omega_{k} a^{\dag}a, \hspace{0.5
cm}k=\pm1.
\end{eqnarray}
The unitary evolution operator relating to Hamiltonian~(\ref{e24})
is $V_{k}(t)=\exp(-i\tilde{H}_{k}t)$. Notice that in Eq.~(\ref{e24})
we have discarded the constant term $C_{k}$.
%%%%%%%%%%%%%%%%%%%%%%%%%%%%%%%%%%%%%%%%%%%%%%%%%%%%%%%%%%%%%%%%%%
\begin{figure}[htp]
\center
\includegraphics[bb=15 15 433 316, width=3.2 in]{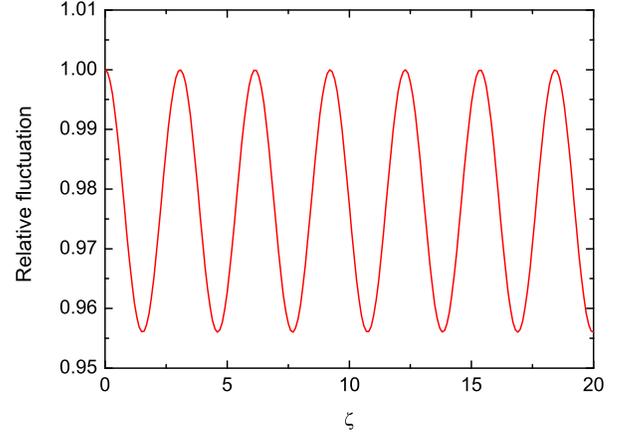}
\caption{(Color online) Plot the relative fluctuation $(\Delta
z)_{-1}^{2}/z_{0}^{2}$ given in Eq.~(\ref{e30}) against the scaled
evolution time $\zeta=\omega t$. Other parameter are set as
$g'/\omega=0.0115$. The relative fluctuation $(\Delta
z)_{-1}^{2}/z_{0}^{2}<1$ means a squeezing of the position $z$ for
the cantilever.}\label{squeezing}
\end{figure}
%%%%%%%%%%%%%%%%%%%%%%%%%%%%%%%%%%%%%%%%%%%%%%%%%%%%%%%%%%%%%%%%%%

For investigation of the squeezing of the cantilever, we assume the
cantilever is initially prepared in a coherent state
$|\alpha\rangle$ and the qubit in states $|\pm\rangle$, where the
coherent amplitude $\alpha$ is assumed to be a real number for
simplicity. After a coherent evolution of a time $t$, the cantilever
evolves into state
\begin{eqnarray}
|\Phi_{k}(t)\rangle=e^{S_{k}}V_{k}(t)e^{-S_{k}}|\alpha\rangle.
\end{eqnarray}
Using the equation $a|\alpha\rangle=\alpha|\alpha\rangle$ and
$|\alpha\rangle=e^{S_{k}}V^{\dagger}_{k}(t)e^{-S_{k}}|\Phi_{k}(t)\rangle$,
we obtain
\begin{eqnarray}
e^{S_{k}}V_{k}(t)e^{-S_{k}}ae^{S_{k}}V_{k}^{\dag}(t)e^{-S_{k}}|\Phi_{k}(t)\rangle=\alpha
|\Phi_{k}(t)\rangle,
\end{eqnarray}
which can be further written as
\begin{eqnarray}
A_{k}(t)|\Phi_{k}(t)\rangle=\alpha |\Phi_{k}(t)\rangle,
\end{eqnarray} with
\begin{eqnarray}
\label{e28}A_{k}(t)&\equiv&e^{S_{k}}V_{k}(t)e^{-S_{k}}ae^{S_{k}}V_{k}^{\dag}(t)e^{-S_{k}}\nonumber\\
&=&[\cos(\Omega_{k} t)+i\cosh(4\lambda_{k})\sin(\Omega_{k}
t)]a\nonumber\\
&&+i\sinh(4\lambda_{k})\sin(\Omega_{k} t)a^{\dag}.
\end{eqnarray}

The evolution of the cantilever follows the eigenstate of the
quasi-excitation operator $A_{k}(t)$ with eigenvalue $\alpha$.
According to Eq.~(\ref{e28}), we can express the operators $a$ and
$a^{\dag}$ in terms of $A_{k}(t)$ and $A^{\dag}_{k}(t)$,
\begin{eqnarray}
\label{e29}a&=&[\cos\Omega_{k}t+i\cosh(4\lambda_{k})\sin(\Omega_{k}t)]A_{k}(t)\nonumber\\
&&+i\sinh(4\lambda_{k})\sin(\Omega_{k}t)A^{\dag}_{k}(t).
\end{eqnarray}
Based on Eq.~(\ref{e29}), the fluctuation $(\Delta
z)_{k}^{2}=\langle z^{2}\rangle_{k}-\langle z\rangle_{k}^{2}$ of the
coordinator $z$ for the cantilever is obtained as
\begin{eqnarray}
\label{e30}(\Delta z)_{k}^{2}(t)
&=&z_{0}^{2}\left[\cosh(4\lambda_{k})-\sinh(4\lambda_{k})\right]^{2}\sin^{2}(\Omega_{k}t)\nonumber\\
&&+\cos^{2}(\Omega_{k}t).
\end{eqnarray}

From Eq.~(\ref{e30}), we find that, only in the case of $k=-1$, the
squeezing of the position $z$ of the cantilever occurs. In
Fig.~\ref{squeezing}, we plot the relative fluctuation $(\Delta
z)_{-1}^{2}/z_{0}^{2}$ as a function of the scaled evolution time
$\zeta=\omega t$. Figure~\ref{squeezing} shows a periodic squeezing
with the scaled evolution time $\zeta$.

\subsection{Dynamical squeezing with dissipation}

In the above subsection, we study the dynamical squeezing for the
ideal case in which there is no dissipation. However, any systems
will couple inevitably with the environment. In this subsection, we
consider the dynamical squeezing of the cantilever by taking the
environment into account. In the presence of an environment, the
evolution of the cantilever is governed by the following quantum
master equation
\begin{equation}
\dot{\rho}=i\left[ \rho \left( t\right),H_{k}\right]
+\mathcal{L}\rho \left( t\right),\label{mastereqsqueezing}
\end{equation}
where the Hamiltonian $H_{k}$ has been given in Eq.~(\ref{e17}), and
the superoperator has been given in Eq.~(\ref{dissipator}). Based on
quantum master equation~(\ref{mastereqsqueezing}), we can obtain the
following equations of motion,
\begin{subequations}
\label{Lequatoins}
\begin{align}
\frac{d}{dt}\langle a\rangle  &=-i\omega _{k}\langle a\rangle
-2ig_{k}\langle a^{\dagger }\rangle -\frac{\gamma
}{2}\langle a\rangle,\\
\frac{d}{dt}\langle a^{\dagger }\rangle  &=i\omega _{k}\langle
a^{\dagger }\rangle +2ig_{k}\langle a\rangle -\frac{\gamma
}{2}\langle a^{\dagger
}\rangle,\\
\frac{d}{dt}\langle a^{2}\rangle &=-( \gamma +2i\omega _{k}) \langle
a^{2}\rangle -4ig_{k}\langle a^{\dagger
}a\rangle -2ig_{k},\\
\frac{d}{dt}\langle a^{\dagger 2}\rangle  &=-( \gamma -2i\omega
_{k})\langle a^{\dagger 2}\rangle
+4ig_{k}\langle a^{\dagger }a\rangle +2ig_{k},\\
\frac{d}{dt}\langle a^{\dagger }a\rangle  &=-\gamma \langle
a^{\dagger }a\rangle -2ig_{k}\langle a^{\dagger 2}\rangle
+2ig_{k}\langle a^{2}\rangle +\gamma \bar{n}_{th}.
\end{align}
\end{subequations}
According to the definition of $z=z_{0}(a+a^{\dagger})$, the
relative fluctuation of the coordinator operator $z$ can be written
as
\begin{eqnarray}
\frac{(\Delta z(t))_{k}^{2}}{z_{0}^{2}} &=&\left(\langle
a^{2}\rangle -\langle a\rangle ^{2}+\langle a^{\dagger 2}\rangle
-\langle a^{\dagger }\rangle ^{2}\right.\nonumber\\&&\left.+2\langle
a^{\dagger }a\rangle -2\langle a^{\dagger }\rangle \langle a\rangle
+1\right).\label{flucdissi}
\end{eqnarray}
Therefore, under a given initial condition, we can obtain the
solutions of Eq.~(\ref{Lequatoins}), and then the evolution of the
fluctuation $(\Delta z)^{2}(t)$ can be obtained. In the following we
suppose the cantilever is initially prepared in a vacuum state
$\vert 0\rangle$.

\begin{figure}[htp]
\center
\includegraphics[bb=6 53 426 595,width=3.2 in]{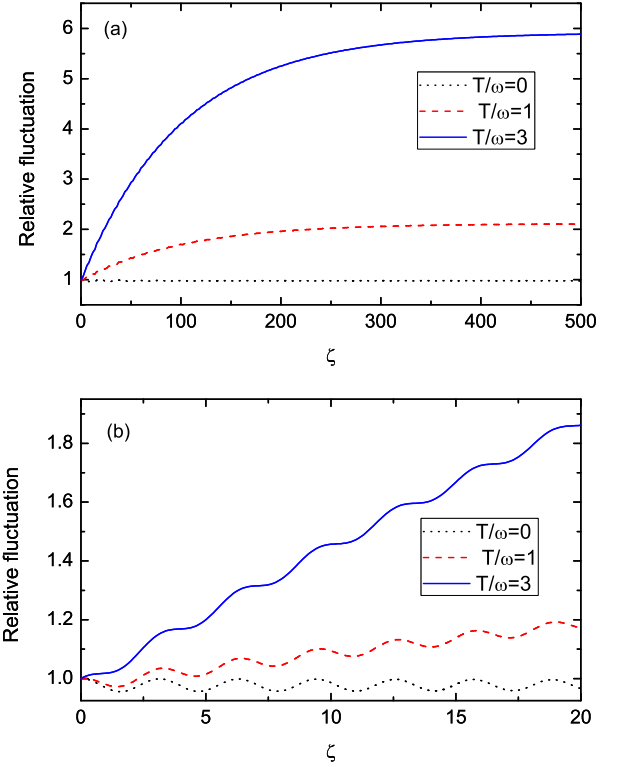}
\caption{(Color online) (a) Plot of the relative fluctuation
$(\Delta z)_{-1}^{2}/z_{0}^{2}$ given in Eq.~(\ref{flucdissi}) vs
the scaled evolution time $\zeta=\omega t$ for different
temperatures: $T/\omega=0$ (dotted line), $T/\omega=1$ (dashed
line), $T/\omega=3$ (solid line). (b) Amplificatory plot of figure
(a) from $\zeta=0$ to $20$. Other parameters are set as
$\gamma/\omega=0.01$, and $g'/\omega=0.0115$.}\label{squeezing2}
\end{figure}

In principle, the solutions of Eq.~(\ref{Lequatoins}) can be
obtained. However, we do not present the solutions here since there
are very complicate. In Fig.~\ref{squeezing2}, we plot the relative
fluctuation $(\Delta z)_{-1}^{2}(t)/z_{0}^{2}$ against the time $t$
for different temperatures.
%%%%%%%%%%%%%%%%%%%%%%%%%%%%%%%%%%%%%%%%%%%%%%%%%%%%%%%%%%
%%%%%%%%%%%%%%%%%%%%%%%%%%%%%%%%%%%%%%%%%%%%%%%%%%%%%%%%%%
It can be seen from Fig.~\ref{squeezing2} that the relative
fluctuation $(\Delta z)_{-1}^{2}/z_{0}^{2}$ evolves gradually
approaching a steady-state value with the increase of the time $t$.
For a short time, the relative fluctuation evolves with some
oscillations. At the same time, the relative fluctuation increases
with the increase of the bath temperature $T$. Therefore, we can
obtain a conclusion that the high temperature $T$ can destroy the
squeezing for the position $z$ of the cantilever.

\begin{figure}[htp]
\center
\includegraphics[bb=25 9 431 308, width=3.2 in]{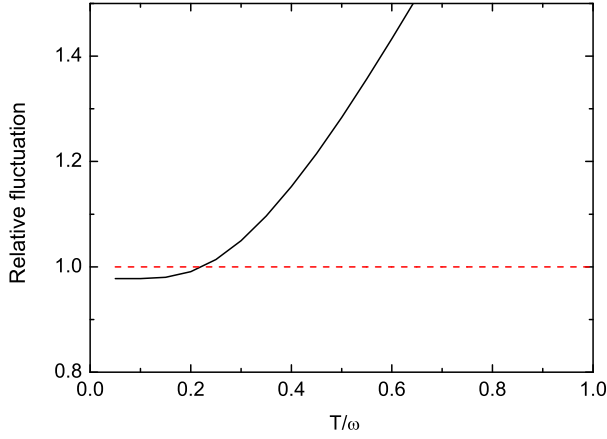}
\caption{(Color online) Plot of the steady-state relative
fluctuation $(\Delta z(\infty))_{-1}^{2}/z_{0}^{2}$ against the
scaled temperature $T/\omega$. Other parameters are set as those in
Fig.~\ref{squeezing2}. Figure shows a transition from squeezing to
nonsqueezing with the increase of the temperature of the
environment.}\label{squeezing3}
\end{figure}

For obtaining the steady-state properties of the squeezing, we
obtain the steady-state solution of Eq.~(\ref{Lequatoins}) as
\begin{subequations}
\begin{align}
\langle a(\infty)\rangle&=\langle
a^{\dagger}(\infty)\rangle=0,\\
\langle
a^{2}(\infty)\rangle&=-\frac{2g_{k}(2\bar{n}_{th}+1)(2\omega_{k}+i\gamma)}{\gamma^{2}+4\omega_{k}^{2}-16g_{k}^{2}},\\
\langle
a^{\dagger2}(\infty)\rangle&=-\frac{2g_{k}(2\bar{n}_{th}+1)(2\omega_{k}-i\gamma)}{\gamma^{2}+4\omega_{k}^{2}-16g_{k}^{2}},\\
\langle
a^{\dagger}a(\infty)\rangle&=\bar{n}_{th}+\frac{8g_{k}^{2}(2\bar{n}_{th}+1)}{\gamma^{2}+4\omega_{k}^{2}-16g_{k}^{2}}.
\end{align}
\end{subequations}
Then the steady-state fluctuation is
\begin{eqnarray}
\frac{(\Delta z(\infty))_{k}^{2}}{z_{0}^{2}}
&=&\frac{(2\bar{n}_{th}+1)[\gamma^{2}+4\omega_{k}(\omega_{k}-2g_{k})]}{\gamma^{2}+4\omega_{k}^{2}-16g_{k}^{2}}.\label{flucdissi}
\end{eqnarray}
In Fig.~\ref{squeezing3}, we plot the steady-state relative
fluctuation $(\Delta z(\infty))_{-1}^{2}/z_{0}^{2}$ as a function of
the temperature $T$. It can be seen from Fig.~\ref{squeezing3} that
the steady-state relative fluctuation increases with the increase of
the temperature $T$. We can see a transition from squeezing to
nonsqueezing when the temperature across a critical temperature
$T_{\textrm{c}}$, which can be obtained from Eq.~(\ref{flucdissi})
for the case of $k=-1$,
\begin{eqnarray}
\frac{T_{\textrm{c}}}{\omega}=\frac{1}{\ln\left(\frac{\gamma^{2}}{4\omega
g'}+\frac{\omega}{g'}+3\right)}.
\end{eqnarray}
According to the parameters, we calculate the critical temperature
in Fig.~\ref{squeezing3} is $T_{\textrm{c}}/\omega=0.222$.
%%%%%%%%%%%%%%%%%%%%%%%%%%%%%%%%%%%%%%%%%%%%%%%%%%%%%%%%%%%%%%%%%%
%%%%%%%%%%%%%%%%%%%%%%%%%%%%%%%%%%%%%%%%%%%%%%%%%%%%%%%%%%%%%%%%%%

\section{\label{parameterestimation}Discussions and conclusions}

We note that the two types of subsystems in our scheme, the
superconducting charge qubit and the cantilever,  have been well
prepared in current experiments. Hence, it is possible to
experimentally realize the scheme proposed in this paper within the
reach of present-day techniques. Here we give a possible estimation
of coupled-system parameters based on these published experimental
parameters of the charge qubit and the cantilevers. The most
important parameters in our model are the two coupling strengths $g$
and $g'$ given in Eqs.~(\ref{e7}) and (\ref{e16}), respectively.
According to the current experimental
conditions~\cite{Sidles1995,Mamin2007}, we take the following
parameters. As an example, we choose a cantilever with a fundamental
frequency $\omega\approx2\pi\times2$ MHz, $\gamma=2\pi\times20$ kHz,
and $z_{0}\approx5\times10^{-13}$ m. A magnetic tip produces a
magnetic gradient of $|\partial B_{z}/\partial z|\approx10^{-7}$ T/m
at an approximate distance $50$ nm above the superconducting loop.
Then we obtian $C=|\partial B_{z}/\partial
z|z_{0}\approx5\times10^{-6}$ T. We choose the area of the
superconducting loop $S\approx10^{-12}$ m$^{2}$, and $E_{J}\approx5$
GHz. Then we get $\phi\approx2.4\pi\times10^{-3}$, which is suitable
for making the approximations of expanding the sine and cosine
functions up to the first and second orders, respectively.
Accordingly, the coupling strengths $g\approx-2\pi\times6$ MHz and
$g'\approx2\pi\times23$ kHz. Notice that the figures in the above
sections are plotted in terms of these parameters.

In conclusion, we have designed a theoretical scheme to realize
tailorable couplings between a cantilever and a superconducting
charge qubit. By choosing proper parameters, both linear and
nonlinear couplings can be achieved. We have also shown how to
generate superposed coherent states and dynamical squeezing in the
cantilever based on the obtained couplings. We have investgated the
influence of the environment on quantum states of the cantilever. It
has been indicated that decoherence induced by the environment can
drive the cantilever from  superposed coherent states  into the
steady coherent state.  When the cantilever is initially in a
coherent state, we have shown that there exists periodic position
squeezing for the cantilever. Especially, it is found that under the
action of the environment the cantilever can evolve from the vacuum
state to a steady state with position squeezing under a critical
temperature of the environment. Therefore, the environment can
induce the steady-state position squeezing of  the cantilever. This
reveals a new mechanism to create the steady-state squeezing and
sheds new light on production of nonclassical effects of the
cantilever. Finally, it should be emphasized that the experimental
realization of the scheme proposed in the present paper deserves
further investigation.

\acknowledgments This work is supported in part by NSFC Grant
No.~10775048, NFRPC Grant No.~2007CB925204, and the Education
Committee of Hunan Province under Grant No. 08W012.

\appendix
\section{\label{appenA}Derivation of quantum master equation~(\ref{transformedmastereq})}

In this Appendix, we give a detailed derivation of the transform
from quantum master equation~(\ref{mastereqT}) to
equation~(\ref{transformedmastereq}). Starting from the quantum
master equation
\begin{eqnarray}
\dot{\rho}&=&i[\rho,\omega a^{\dagger}a+g_{s}(a^{\dagger
}+a)]\nonumber\\ &&+\frac{\gamma }{2}(\bar{n}+1)(2a\rho a^{\dagger
}-a^{\dagger }a\rho -\rho
a^{\dagger }a) \nonumber\\
&&+\frac{\gamma }{2}\bar{n}(2a^{\dagger }\rho a-aa^{\dagger }\rho
-\rho aa^{\dagger }),\label{appenA1}
\end{eqnarray}
we first make a rotation transform,
\begin{eqnarray}
\rho^{\left( 1\right) }=R^{\dagger }\left( \theta \right) \rho
R\left( \theta \right), \hspace{0.5 cm}R\left( \theta
\right)=e^{-i\theta a^{\dagger }a},
\end{eqnarray}
then we obtain $\rho  =R\left( \theta \right) \rho^{\left( 1\right)
} R^{\dagger }\left( \theta \right)$ and
\begin{eqnarray}
\dot{\rho} &=&-i\dot{\theta}a^{\dagger }aR\left( \theta \right)
\rho^{\left( 1\right) }R^{\dagger }\left( \theta \right) +R\left(
\theta \right) \dot{\rho}^{\left( 1\right) } R^{\dagger }\left(
\theta \right)\nonumber\\&& +i\dot{\theta}R\left( \theta \right)
\rho^{\left( 1\right) }  a^{\dagger }aR^{\dagger }\left( \theta
\right).\label{paudots}
\end{eqnarray}
Substitution of Eq.~(\ref{paudots}) into quantum master
equation~(\ref{appenA1}) leads to
\begin{eqnarray}
\dot{\rho}^{\left( 1\right) }  &=&i(\omega -\dot{ \theta})
\rho^{\left( 1\right) }a^{\dagger }a-i( \omega -\dot{\theta})
a^{\dagger }a\rho^{\left(
1\right) } \nonumber \\
&&+ig_{s}\rho^{\left( 1\right) }\left( a^{\dagger }e^{i\theta
}+ae^{-i\theta }\right) -ig_{s}\left( a^{\dagger }e^{i\theta
}+ae^{-i\theta }\right) \rho^{\left( 1\right) }\nonumber\\
&&+\frac{\gamma }{2}\left( \bar{n}+1\right)\left(2a\rho^{(1)}
a^{\dagger }-a^{\dagger }a\rho^{(1)}-\rho^{(1)}
a^{\dagger }a\right)\nonumber \\
&&+\frac{\gamma }{2}\bar{n}\left(2a^{\dagger }\rho^{(1)}
a-aa^{\dagger }\rho^{(1)}-\rho^{(1)}aa^{\dagger }\right),
\end{eqnarray}
where we have used the relations
\begin{subequations}
\begin{align}
\left[ R\left( \theta \right) ,a^{\dagger }a\right] &=\left[
R^{\dagger
}\left( \theta \right) ,a^{\dagger }a\right] =0,\\
R^{\dagger }\left( \theta \right) a^{\dagger }R\left( \theta \right)
&=e^{i\theta a^{\dagger }a}a^{\dagger }e^{-i\theta a^{\dagger
}a}=a^{\dagger }e^{i\theta },\\
R^{\dagger }\left( \theta \right) aR\left( \theta \right)
&=e^{i\theta a^{\dagger }a}ae^{-i\theta a^{\dagger
}a}=ae^{-i\theta}.
\end{align}
\end{subequations}
We choose a proper $\theta$ to ensure $\omega -\dot{\theta}=0$.
Under the initial condition $\theta(0)=0$, we get
\begin{eqnarray}
\theta(t) =\omega t.
\end{eqnarray}
Then we obtain
\begin{eqnarray}
\dot{\rho}^{(1) }&=&ig_{s}\left[\rho^{(1)} ,a^{\dagger }e^{i\omega
t}+ae^{-i\omega t}\right] \nonumber\\
&&+\frac{\gamma }{2}(\bar{n}+1)\left(2a\rho ^{(1)}a^{\dagger
}-a^{\dagger }a\rho^{(1)
}-\rho^{(1)}a^{\dagger }a\right) \nonumber\\
&&+\frac{\gamma }{2}\bar{n}\left( 2a^{\dagger }\rho^{(1)
}a-aa^{\dagger }\rho^{(1)}-\rho^{(1) }aa^{\dagger
}\right).\label{rottransformed}
\end{eqnarray}
Since the first term at the right-hand side of
Eq.~(\ref{rottransformed}) is a driving term, in the following, we
make a displacement transform
\begin{eqnarray}
\rho^{\left( 2\right)} &=&D^{\dagger }\left( \alpha \right)
\rho^{\left( 1\right)}D\left( \alpha \right),\hspace{0.5 cm} D\left(
\alpha \right)=e^{\alpha a^{\dagger }-\alpha ^{\ast }a},
\end{eqnarray}
then we obtain $\rho^{\left( 1\right) }=D\left( \alpha \right)
\rho^{\left( 2\right) }  D^{\dagger }\left( \alpha \right)$ and
\begin{eqnarray}
\dot{\rho}^{\left( 1\right) }  &=&\dot{D}\left( \alpha \right)
\rho^{\left( 2\right) }  D^{\dagger }\left( \alpha \right) +D\left(
\alpha \right) \dot{\rho}^{\left( 2\right) } D^{\dagger }\left(
\alpha \right)\nonumber\\&& +D\left( \alpha \right) \rho^{\left(
2\right) }\left( t\right) \dot{D}^{\dagger }\left( \alpha
\right).\label{paudot}
\end{eqnarray}
Here we need to calculate the expressions for $\dot{D}\left( \alpha
\right)$ and $\dot{D}^{\dagger }\left( \alpha \right)$. Making use
of $D\left( \alpha \right) =e^{-\frac{%
\left\vert \alpha \right\vert ^{2}}{2}}e^{\alpha a^{\dagger
}}e^{-\alpha ^{\ast }a}$, we have
\begin{eqnarray}
\dot{D}\left( \alpha \right) &=&\frac{d}{dt}\left( e^{-\frac{\left\vert \alpha \right\vert ^{2}}{2}%
}\right) e^{\alpha a^{\dagger }}e^{-\alpha ^{\ast
}a}+e^{-\frac{\left\vert \alpha \right\vert
^{2}}{2}}\frac{d}{dt}\left( e^{\alpha a^{\dagger
}}\right) e^{-\alpha ^{\ast }a}\nonumber\\
&&+e^{-\frac{\left\vert \alpha \right\vert ^{2}%
}{2}}e^{\alpha a^{\dagger }}\frac{d}{dt}\left( e^{-\alpha ^{\ast
}a}\right)
\nonumber\\
&=&-\frac{1}{2}( \dot{\alpha}\alpha ^{\ast }+\alpha
\dot{\alpha}^{\ast }) D(\alpha) +\dot{\alpha}D(\alpha) e^{\alpha
^{\ast }a}a^{\dagger
}e^{-\alpha ^{\ast }a}\nonumber\\
&&-\dot{\alpha}^{\ast
}D(\alpha) a \nonumber\\
&=&-\frac{1}{2}\left(\alpha \dot{\alpha}^{\ast }-\dot{\alpha}\alpha
^{\ast }\right)D(\alpha)+D(\alpha) (\dot{\alpha} a^{\dagger
}-\dot{\alpha}^{\ast
}a),\label{Ddot}\nonumber\\
\end{eqnarray}
where we have used the formula $e^{\alpha ^{\ast }a}a^{\dagger
}e^{-\alpha ^{\ast }a}=a^{\dagger }+\alpha ^{\ast }$. Similarly, we
obtain
\begin{eqnarray}
\dot{D}^{\dagger }(\alpha) &=&-\frac{1}{2}(\dot{\alpha}\alpha
^{\ast}-\alpha \dot{\alpha}^{\ast }) D^{\dagger
}(\alpha)+(\dot{\alpha}^{\ast}a-
\dot{\alpha}a^{\dagger})D^{\dagger }(\alpha).\label{Ddaggerdot}\nonumber\\
\end{eqnarray}

Substitution of Eqs.~(\ref{paudot}), (\ref{Ddot}), and
(\ref{Ddaggerdot}) into Eq.~(\ref{rottransformed}), we obtain
\begin{eqnarray}
\dot{\rho}^{(2)}&=&\frac{\gamma }{2}(\bar{n}+1)\left( 2a\rho^{(2)}
a^{\dagger}-a^{\dagger }a\rho^{(2)}-\rho^{(2)}a^{\dagger}a\right)\nonumber\\
&&+\frac{\gamma }{2}\bar{n}\left(2a^{\dagger }\rho^{(2)}
a-aa^{\dagger
}\rho^{(2)}-\rho^{(2)}  aa^{\dagger}\right)\nonumber\\
&&+\left(\dot{\alpha}^{\ast }-ig_{s}e^{-i\omega t}+\frac{\gamma
}{2}\alpha
^{\ast }\right)\left( a\rho^{(2)}(t)-\rho_{s}^{(2)}a\right)\nonumber\\
&&-\left(\dot{\alpha}+ig_{s}e^{i\omega t}+\frac{\gamma
}{2}\alpha\right)\left( a^{\dagger }\rho^{(2)}-\rho^{(2)}a^{\dagger
}\right).
\end{eqnarray}

Therefore, if we choose a proper $\alpha $ such that
\begin{eqnarray}
\dot{\alpha}+ig_{s}e^{i\omega t}+\frac{\gamma }{2}\alpha
&=&0,\label{alphaequation}
\end{eqnarray}
then we obtain the quantum master
equation~(\ref{transformedmastereq}). The solution of
Eq.~(\ref{alphaequation}) can be obtained as
\begin{eqnarray}
\alpha(t)&=&\frac{ig_{s}}{\frac{\gamma }{2}+i\omega }\left(
e^{-\frac{\gamma t}{2}}-e^{i\omega t}\right)
\end{eqnarray}
under the initial condition $\alpha(0)=0$.

\section{\label{appendixb}Derivation of Eq.~(\ref{stateatt})}

In this Appendix, we derive in detail the evolution of the
cantilever governed by quantum master equation~(\ref{mastereqT}).
For an initial state $\rho(0)$, the state of the cantilever at time
$t$ can be obtained through the processes
$\rho(0)\rightarrow\rho^{(2)}(0)\rightarrow\rho^{(2)}(t)\rightarrow\rho(t)$.
The relationship between states $\rho(0)$ and $\rho^{(2)}(0)$ is
\begin{eqnarray}
\rho^{(2)}(0)&=&D^{\dagger}(\alpha(0))R^{\dagger}(\theta(0))\rho(0)R(\theta(0))D(\alpha(0))\nonumber\\
&=&\rho(0).
\end{eqnarray}
For the initial state
\begin{equation}
\vert \psi(0)\rangle =\mathcal{N}_{\varphi}(\vert \beta\rangle
+e^{i\varphi}\vert -\beta\rangle),
\end{equation}
we have
\begin{eqnarray}
\rho^{(2)}(0)&=&\mathcal{N}_{\varphi}^{2}(\vert \beta\rangle
+e^{i\varphi }\vert -\beta\rangle) (\langle \beta\vert
+e^{-i\varphi}\langle -\beta\vert ).\label{eqb3}
\end{eqnarray}

The evolution process from state $\rho ^{\left( 2\right) }\left(
0\right) $ to $\rho ^{\left( 2\right) }\left( t\right) $ is governed
by the quantum master equation
\begin{eqnarray}
\dot{\rho}^{(2)}(t)&=&\frac{\gamma }{2}(\bar{n}+1)\left(2a\rho
^{\left(
2\right) }(t)a^{\dagger }-a^{\dagger }a\rho ^{(2)}(t)-\rho^{(2) }(t)a^{\dagger }a\right)\nonumber\\
&&+\frac{\gamma }{2}\bar{n}\left(2a^{\dagger}\rho^{\left(
2\right)}a-aa^{\dagger }\rho ^{\left( 2\right) }\left( t\right)
-\rho^{\left(2\right)}(t) aa^{\dagger }\right).
\end{eqnarray}
In principle, the above master equation can be solved by the
superoperator method~\cite{Burnett}. However, for simplicity, we
only consider the zero temperature case in the following. At zero
temperature, the master equation reduces to
\begin{eqnarray}
\dot{\rho}^{(2)}&=&\frac{\gamma }{2}\left(2a\rho ^{(2)}(t)
a^{\dagger }-a^{\dagger }a\rho^{(2)}(t) -\rho^{(2)}(t) a^{\dagger
}a\right).
\end{eqnarray}
Denoting the two superoperators
\begin{subequations}
\begin{align}
J\rho^{(2)}(t)&=a\rho^{(2)}(t) a^{\dagger },\\
L\rho^{(2)}(t)&=-\frac{1}{2}\left(a^{\dagger }a\rho^{(2)}(t)
+\rho^{(2)}(t)a^{\dagger }a\right),
\end{align}
\end{subequations}
then the map from initial state $\rho^{(2)}(0)$ to $\rho^{(2)}(t)$
at time $t$ is determined by
\begin{equation}
\rho^{(2)}(t)=\exp(\gamma tL)\exp[(1-\exp(-\gamma
t))J]\rho^{(2)}(0).
\end{equation}

According to the initial state given in Eq.~(\ref{eqb3}), we have
\begin{eqnarray}
\rho ^{\left( 2\right) }\left(
t\right)&=&\mathcal{N}_{\varphi}^{2}\left( \left\vert \beta e^{-
\frac{\gamma t}{2}}\right\rangle \left\langle \beta e^{-\frac{\gamma
t}{2} }\right\vert +e^{-i\varphi }e^{\Delta_{r}}\left\vert \beta
e^{-\frac{\gamma t}{2}}\right\rangle \left\langle -\beta
e^{-\frac{\gamma t}{
2}}\right\vert\right.\nonumber\\&&\left.+e^{i\varphi
}e^{\Delta_{r}}\left\vert -\beta e^{-\frac{\gamma
t}{2}}\right\rangle \left\langle \beta e^{-\frac{
\gamma t}{2}}\right\vert +\left\vert -\beta e^{-\frac{\gamma t}{2}%
}\right\rangle \left\langle -\beta e^{-\frac{\gamma
t}{2}}\right\vert \right),\nonumber\\\label{B10}
\end{eqnarray}
where we introduce the parameter $\Delta _{r}=2\vert\beta\vert
^{2}(e^{-\gamma t}-1)$. During the derivation of the above
equation~(\ref{B10}), we have used the formulas
\begin{subequations}
\begin{align}
\exp[xJ]|r\rangle\langle s|&=\exp[ x r s^{\ast
}]\vert r\rangle\langle s\vert,\\
\exp(\gamma tL)\vert r \rangle \langle s \vert &=\exp[-(\vert
r\vert^{2}+\vert s\vert^{2})(
1-e^{-\gamma t})/2]\nonumber\\
&\times \left\vert r e^{-\frac{\gamma t}{2}}\right\rangle
\left\langle s e^{-\frac{\gamma t}{2}}\right\vert,
\end{align}
\end{subequations}
for coherent states $|r\rangle$ and $|s\rangle$.

For obtaining the state in the Schr\"{o}dinger picture at time $t$,
we use the relation
$\rho^{(1)}(t)=D(\alpha(t))\rho^{(2)}(t)D^{\dagger}(\alpha(t))$ to
obtain
\begin{eqnarray}
&&D\left(\alpha \left( t\right) \right) \rho ^{\left( 2\right)
}\left( t\right) D^{\dagger }\left(\alpha \left( t\right)
\right)\nonumber\\
 =&&\mathcal{N}_{\varphi}^{2}\left(
\left\vert \alpha \left( t\right) +\beta e^{-\frac{\gamma t}{2}
}\right\rangle \left\langle \alpha \left( t\right) +\beta
e^{-\frac{\gamma t }{2}}\right\vert
\right.\nonumber\\&&\left.+\left\vert \alpha \left( t\right) -\beta
e^{-\frac{\gamma t }{2}}\right\rangle \left\langle \alpha \left(
t\right) -\beta e^{-\frac{ \gamma t}{2}}\right\vert
\right.\nonumber\\&&\left.+e^{-i\varphi }e^{\Delta _{r}+i\Delta
_{i}}\left\vert \alpha \left( t\right) +\beta e^{-\frac{\gamma t}{2}
}\right\rangle \left\langle \alpha \left( t\right) -\beta
e^{-\frac{\gamma t }{2}}\right\vert
\right.\nonumber\\&&\left.+e^{i\varphi }e^{\Delta _{r}-i\Delta
_{i}}\left\vert \alpha \left( t\right) -\beta e^{-\frac{\gamma
t}{2}}\right\rangle \left\langle \alpha \left( t\right) +\beta
e^{-\frac{\gamma t}{2}}\right\vert \right)
\end{eqnarray}
where we introduce the parameter $\Delta _{i}=2\textrm{Im}(
\alpha(t) \beta ^{\ast }\exp(-\gamma t/2))$. Here we have used the
formula
\begin{eqnarray}
D(\alpha)\vert\beta\rangle =e^{i\textrm{Im}(\alpha \beta ^{\ast
})}\vert\alpha+\beta\rangle
\end{eqnarray}
for coherent state $\vert \beta\rangle$. And then we use the
relation $\rho(t)=R(\theta)\rho^{(1)}(t)R^{\dagger}(\theta)$ to
obtain the state
\begin{eqnarray}
\rho \left( t\right) &=&\mathcal{N}_{\varphi}^{2}\left( \left\vert
\beta _{+}\right\rangle \left\langle \beta _{+}\right\vert
+e^{-i\varphi }e^{\Delta _{r}+i\Delta _{i}}\left\vert \beta
_{+}\right\rangle \left\langle \beta
_{-}\right\vert \right.\nonumber\\
&&\left.+e^{i\varphi }e^{\Delta _{r}-i\Delta _{i}} \left\vert \beta
_{-}\right\rangle \left\langle \beta _{+}\right\vert +\left\vert
\beta _{-}\right\rangle \left\langle \beta _{-}\right\vert\right),
\end{eqnarray}
where we introduce the following two parameters
\begin{eqnarray}
\beta_{\pm}=\left(\alpha(t)\pm\beta e^{-\frac{\gamma t}{2}}\right)
e^{-i\omega t}.\label{parameters}
\end{eqnarray}

%%%%%%%%%%%%%%%%%%%%%%%%%%%%%%%%%%%%%%%%%%%%%%%%%%%%%%%%%%%%%%%%%%%%%%%%%%%%%%%%%%%%%%%%%%%%%%%%%%%%%%%%%%%%%%%%%%%%%%%%%%%%

%%%%%%%%%%%%%%%%%%%%%%%%%%%%%%%%%%%%%%%%%%%%%%%%%%%%%%%%%%%%%%%%%%%%%%%%%%%%%%%%%%%%%%%%%%%%%%%%%%%%%%%%

\end{document}